\documentclass{siamart0216}
\pdfoutput=1

\usepackage{lipsum}
\usepackage{amsfonts}
\usepackage{graphicx}
\usepackage{algorithmic}
\ifpdf
  \DeclareGraphicsExtensions{.eps,.pdf,.png,.jpg}
\else
  \DeclareGraphicsExtensions{.eps}
\fi

\newcommand{\TheTitle}{Fast Calculation of the Knowledge Gradient for Optimization of Deterministic Engineering Simulations} 
\newcommand{\TheAuthors}{J. van der Herten, I. Couckuyt, D. Deschrijver and T. Dhaene}

\headers{Knowledge Gradient for Deterministic Problems}{\TheAuthors}
\title{{\TheTitle}}

\author{
  Joachim van der Herten\thanks{Internet Based Communication Networks and Services (IBCN), Technologiepark 15, 9052 Ghent, Belgium
    (\email{joachim.vanderherten@intec.ugent.be}, \email{ivo.couckuyt@intec.ugent.be}, \email{dirk.deschrijver@intec.ugent.be}, \email{tom.dhaene@intec.ugent.be}, \url{http://sumo.intec.ugent.be}).}
  \and
  Ivo Couckuyt\footnotemark[1] \thanks{Ivo Couckuyt is a post-doctoral research fellow of the Research Foundation Flanders (FWO).}
  \and
  Dirk Deschrijver\footnotemark[1]
  \and
  Tom Dhaene\footnotemark[1]
}

\usepackage{amsopn}

\ifpdf
\hypersetup{
  pdftitle={\TheTitle},
  pdfauthor={\TheAuthors}
}
\fi

\usepackage[caption=false]{subfig}
\usepackage{commath}
\usepackage{siunitx}

\DeclareMathOperator*{\argmax}{arg\,max}

\newcommand{\specialcell}[2][c]{%
  \begin{tabular}[c]{@{}#1@{}}#2\end{tabular}}
  
\def\bfx{\mbox{\boldmath$x$}}

\def\bfu{\mbox{\boldmath$u$}}

\def\bfalpha{\mbox{\boldmath$\alpha$}}
\def\bftheta{\mbox{\boldmath$\theta$}}

\begin{document}

\maketitle

\begin{abstract}
A novel efficient method for computing the Knowledge-Gradient policy for Continuous Parameters (KGCP) for deterministic optimization is derived. The differences with Expected Improvement (EI), a popular choice for Bayesian optimization of deterministic engineering simulations, are explored. Both policies and the Upper Confidence Bound (UCB) policy are compared on a number of benchmark functions including a problem from structural dynamics. It is empirically shown that KGCP has similar performance as the EI policy for many problems, but has better convergence properties for complex (multi-modal) optimization problems as it emphasizes more on exploration when the model is confident about the shape of optimal regions. In addition, the relationship between Maximum Likelihood Estimation (MLE) and slice sampling for estimation of the hyperparameters of the underlying models, and the complexity of the problem at hand, is studied.
\end{abstract}



\section{Introduction}
As building many prototypes and performing real-life experiments is costly, engineers have adopted the concepts of virtual prototyping and computer aided design (CAD) since long. After specifying a set of values for the input parameters (design of experiments), multiple experiments on the complex input-output systems are performed virtually by means of computer simulations resulting in cost savings and a shorter time-to-market.

Over the years, the accuracy of the available simulation software has improved significantly allowing simulation of systems at a finer level of detail. This evolution opens up usage of simulations for increasingly complex problems, but also increases the associated computational cost tremendously. Some high-fidelity simulations are known to require days or even weeks of runtime for a single evaluation \cite{Goethals2012}. This makes their use infeasible for evaluation-intensive analysis such as parameter exploration, sensitivity analysis or optimization. This sparked the development of surrogate modeling or metamodeling: essentially these are predictive models used specifically to approximate the behavioral response of engineering systems. The simulator responses should be approximated accurately using a minimum number of evaluations, while still keeping the computational cost acceptable \cite{Gorissen2010b}. 

Surrogate models can either be used as a global approximation that can replace the simulator, but can also be used to guide an optimization process. Most known is the Efficient Global Optimization (EGO) methodology \cite{Movckus1975, Jones1998} which sequentially picks the next evaluation by optimizing the Expected Improvement (EI) policy using Kriging models under the assumption that the response of the simulation is deterministic: the output of the simulation is considered to be noise-free, the only form of noise encountered is negligibly small, e.g., in the order of the machine epsilon. The EI policy essentially observes the difference between the expectancy on the prediction mean of the next intermediate model given an arbitrary unobserved evaluation and the current best observed value. It has been shown that the concept can be extended to multi-objective optimization problems by aggregating the responses of several objectives \cite{Knowles2006,Hernandez2015} or by allowing different interpretations of improvement \cite{Emmerich2011,Couckuyt2014b}. This approach has been applied successfully for optimization of several engineering applications and quickly leads to satisfying results. Because the objective function is expensive to compute, spending some computation time to decide the next evaluation intelligently is justified. The EI policy, amongst others, can be combined with different (Bayesian) models to optimize any real-valued (typically expensive) objective function: this is often referred to as Bayesian Optimization (BO) which has lately become increasingly popular for hyperparameter optimization \cite{Bergstra2011,Snoek2012,Gardner2014}.

A different policy for Bayesian optimization is the Upper Confidence Bound (UCB) \cite{Cox1997}. Although the concept of the UCB policy is quite straightforward, it features some strong theoretical guarantees \cite{Freitas2012}. Another policy for Bayesian optimization with a discrete set of candidate evaluations in the presence of uncertainty on the obtained response is the Knowledge-Gradient for Correlated Beliefs (KGCB) \cite{Frazier2009}. This policy has been extended to continuous parameter intervals known as the Knowledge Gradient for Continuous Parameters (KGCP) \cite{Scott2011}. The relationship to, and differences with EI have been discussed in \cite{Powell2012}, although sometimes the two policies are mixed up \cite{Bull2011}. Results obtained on problems involving uncertain responses (i.e., noise) pointed out an advantage of KGCP over EGO and Sequential Kriging Optimization (SKO) \cite{Huang2006,Picheny2013} which both use EI, the latter including a correction term to account for the belief that the unknown next point to be evaluated also has noise associated with it. However for deterministic problems (as often encountered in physics-based engineering simulation) the additional complexity of computing the KGCP has been disproportional to the advantage in terms of evaluations. 

In short, this article addresses the following:
\begin{itemize}
\item A novel closed form for computation of the KGCP for deterministic problems is derived, and it is shown that the KGCP has now similar (computational) complexity as EI.
\item The relationship between EI and KGCP is studied. From the formulation it can be observed the KGCP has more confidence in the underlying intermediate model, as compared to EI.
\item The KGCP is compared with EI and UCB on several deterministic functions and a real-life 10D structural dynamics optimization problem from engineering. 
\item In addition, the use of slice sampling and MLE for selecting the hyperparameters of the underlying Kriging model is compared empirically on all test problems.
\end{itemize}
Although most recent research on Bayesian optimization focusses on stochastic optimization, deterministic problems arise frequently in various research application domains and are frequently solved efficiently by (Bayesian) machine learning methods. Machine learning-­based engineering is a prime example. Here expensive deterministic simulations are optimized using machine learning methods. The EI policy has long been a popular choice for this task. Optimization of deterministic engineering problems is the point of view of this article, though our approach to compute the KGCP is general and can be applied for any (expensive) deterministic optimization problem. 

\Cref{sec:formalism} addresses the optimization problem formally and introduces the model used by both policies in this contribution. \Cref{sec:kg} reviews the Knowledge-Gradient and develops a closed form to compute the KGCP for deterministic problems. In \cref{sec:exp} the KGCP is compared to EI and UCB on several deterministic problems.

\section{Formalism}
\label{sec:formalism}
Given the following global optimization problem:
\begin{equation}
\label{eq:globaloptim}
\underset{\bfx \in \mathcal{X}}{\argmax} ~ \mu(\bfx),
\end{equation}
for an unknown function $\mu: \mathbb{R}^d \rightarrow \mathbb{R}$ corresponding to the simulator. The vector $\bfx$ is referred to as \textit{decision} and is part of the set of feasible decisions $\mathcal{X} \subset \mathbb{R}^d$. We can obtain the observation $y_n$ by evaluating $\mu$ for a decision $\bfx_n$. Assuming the observation is not exact and subject to uncertainty, the distribution of $y_n$ is centered around the true response with variance given by $\lambda: \mathbb{R}^d \rightarrow \mathbb{R}$. We focus on deterministic problems, which implies $\lambda(x_n) = 0$. The goal is a sequential set of decisions $\bfx_i, i = 0,..., N-1$ ($N$ being the total amount of allowed evaluations) using a sampling policy so that the probability of identifying the optimal solution to \cref{eq:globaloptim} using only a limited number of observations is maximal. Because evaluating $\mu$ is expensive, additional computational effort to determine these decisions is justified.

\subsection{Kriging interpolation}
We proceed by modeling the obtained information on $\mu$ with a Kriging model \cite{Couckuyt2014}. Given $n < N$ observations $y_i$ corresponding to decisions $\{\bfx_0,...,\bfx_{n-1}\}$, we construct a model $\mathcal{M}^n$ with prediction mean and variance $\mu^n$ and $s^n$ respectively, which combines a regression model and Gaussian process with mean 0, variance $\sigma^2$ and correlation matrix $\Psi$ interpolating the residual. 

Given a set of basis functions $B = \{b_1,...,b_p\}$ and a correlation function $\psi$, 
$F$ and $\Psi$ represent the regression and correlation matrix respectively:
\begin{equation*}
F = \begin{bmatrix}
    b_1(\bfx_0)       & \hdots & b_p(\bfx_{0}) \\
    \vdots       & \ddots & \vdots \\
    b_1(\bfx_{n-1})       & \hdots & b_p(\bfx_{n-1})
\end{bmatrix}, ~ \Psi = \begin{bmatrix}
    \psi(\bfx_0, \bfx_0)       & \hdots & \psi(\bfx_{0}, \bfx_{n-1}) \\
    \vdots       & \ddots & \vdots \\
    \psi(\bfx_{n-1}, \bfx_0)       & \hdots & \psi(\bfx_{n-1}, \bfx_{n-1})
\end{bmatrix}.
\end{equation*}
The vector $m(\bfx) = \left( b_1(\bfx),...,b_p(\bfx) \right)$ corresponds to the arbitrary decision $\bfx$ evaluated on all basisfunctions, whereas $r(\bfx) = \left( \psi(\bfx, \bfx_1),...,\psi(\bfx, \bfx_n) \right)$. The regression coefficients $\bfalpha$ can be obtained by solving the Generalized Least Squares problem:
\begin{equation*}
\bfalpha = \left( F^T \Psi^{-1} F \right)^{-1} F^T \Psi^{-1} y. 
\end{equation*} 
The process variance is given by $\sigma^2 = \frac{1}{n} (y-F\bfalpha)^T \Psi^{-1} (y-F\bfalpha)$. For this model, the prediction mean and variance are given by
\begin{subequations}
\begin{align}
\mu^n(\bfx) &= m(\bfx) \bfalpha + r(\bfx) \Psi^{-1} (y - F \bfalpha), \\
s^n(\bfx) &= \sigma^2 \left( 1 - r(\bfx) \Psi^{-1} r(\bfx)^T \frac{1-F^T\Psi^{-1}r(\bfx)^T}{F^T \Psi^{-1} F}\right).
\end{align}
\end{subequations}
The gradients for prediction mean and variance for the model can also be computed analytically:
\begin{subequations}
\begin{align}
\frac{\dif \mu^n(\bfx)}{\bfx} &= J_{F} \bfalpha + J_{\Psi} (y - F \bfalpha),\\
\frac{\dif s^n(\bfx)}{\bfx} &= 2\sigma^2 \left( \left( F^{T} \Psi^{-1} F\right)^{-1} \left(F^{T} \Psi^{-1} r(\bfx)^T - J_{F} \bfalpha \right) - \Psi^{-1} r(\bfx)^{T} \right), \\
J_{F} &= \begin{bmatrix}
    \frac{\dif b_1(\bfx)}{\dif x_1}       & \hdots & \frac{\dif b_p(\bfx)}{\dif x_1} \\
    \vdots       & \ddots & \vdots \\
    \frac{\dif b_1(\bfx)}{\dif x_n}       & \hdots & \frac{\dif b_p(\bfx)}{\dif x_n}
\end{bmatrix}, ~ J_{\Psi} = \begin{bmatrix}
    \frac{\dif \psi(\bfx_1, \bfx)}{\dif x_1}       & \hdots & \frac{\dif \psi(\bfx_n, \bfx)}{\dif x_1} \\
    \vdots       & \ddots & \vdots \\
    \frac{\dif \psi(\bfx_1, \bfx)}{\dif x_p}       & \hdots & \frac{\dif \psi(\bfx_n, \bfx)}{\dif x_n}
\end{bmatrix}.
\end{align}
\end{subequations}
Under this formulation the model interpolates the training data. By adding a constant to the diagonal of $\Psi$ the model becomes a regression model and is referred to as Stochastic Kriging in operational research. Note that in machine learning, policies are most often based on a Gaussian process with noise. Kriging can be considered a special case of a Gaussian process where the regression function coefficients are estimated by Generalized Least Squares. We chose the Kriging formulation instead, because it is a common choice in engineering simulation optimization (as it is the model type used in the EGO formulation \cite{Jones1998}).

\subsection{Basis and Correlation functions}
\label{sec:hpoptim}
Ordinary kriging is the common choice and includes only the constant regression function: $b_0(\bfx) = 1$. Specifying a correlation function $\psi$ defines the correlation matrix $\Psi$. We selected the Mat\'{e}rn 5/2 correlation function \cite{Stein2012} for engineering problems, because the popular Gaussian correlation function assumes an unrealistic smoothness of the underlying response \cite{Snoek2012}. The Mat\'{e}rn 5/2 correlation function requires only twice differentiability which fits many engineering problems,  including the applications tackled in this work. Assuming two decisions $\bfx_i$ and $\bfx_j$ we get
\begin{align*}
\psi(\bfx_i,\bfx_j) &= \left( 1 + \sqrt{5} l + \frac{5l^2}{3} \right) \exp \left( -5 \sqrt{l} \right) \\
l & = \sqrt{ (\bfx_i - \bfx_j)^T \text{diag}\left( \bftheta \right) (\bfx_i - \bfx_j) }.
\end{align*}
The choice of the hyperparameter vector\footnote{We consider \textit{anisotropic} correlation functions with a hyperparameter for each dimension. In the \textit{isotropic} case $\theta \in \mathbb{R}^1$.} $\bftheta \in \mathbb{R}^d$ is crucial to obtain a meaningful prediction. The hyperparameter vector is typically identified using Maximum Likelihood Estimation (MLE). Several variants of the likelihood are available, most commonly used is the negative concentrated log-likelihood:
\begin{equation*}
-\text{ln} \left( \mathcal{L} \right) = -\frac{1}{2} \left( n ~ \text{ln}(\sigma^2) + \text{ln}(|\Psi|) \right).
\end{equation*}
However it was reported the MLE solution can result in a biased prediction variance \cite{Kleijnen2006}. As the uncertainty expressed by the prediction variance is crucial in Bayesian optimization, the uncertainty about the model parameters should be incorporated which cannot be accomplished using a single parameter estimate. Hence we also consider slice sampling the likelihood as proposed previously \cite{Murray2010, Snoek2012} and generate a set of hyperparameter vectors $\{\bftheta_i\}_{i=1}^h$. The prediction mean and variance are obtained by averaging the responses using all obtained $\bftheta$ vectors. Similarly for computing the sampling policies, their responses are computed by averaging the scores obtained using all $\bftheta$ vectors instead of using the averaged prediction mean and variance and computing the score only once.

\section{Knowledge-Gradient Policy}
\label{sec:kg}
The knowledge-gradient policy was described in \cite{Frazier2009} for optimization over a discrete decision domain $\mathcal{X}$. At iteration $n$, we have obtained $n$ observations $\{y_0,...y_{n-1}\}$ corresponding to $n$ decisions $\{\bfx_0,...,\bfx_{n-1}\}, \bfx_i \in \mathcal{X}$. The information gained from measuring $\bfx \in \mathcal{X}$ is defined as the knowledge-gradient: 
\begin{equation}
\label{eq:kgcb}
\nu^{\mbox{KG},n}(\bfx) = \mathbb{E} \left[ \underset{\bfu \in \mathcal{X}}{\max} ~ \mu^{n+1}(\bfu)| \bfx_n = \bfx \right]-\underset{\bfu\in \mathcal{X}}{\max} ~ \mu^{n}(\bfu).
\end{equation}
The next sampling decision $\bfx_n$ is chosen as the maximum over the knowledge-gradient,
\begin{equation}
\label{eq:kgcpoptimization}
\bfx^n \in \underset{\bfx \in \mathcal{X}}{\argmax} ~ \nu^{\mbox{KG},n}(\bfx).
\end{equation}
This corresponds to the knowledge-gradient policy. An algorithm to solve \cref{eq:kgcb} was formulated in \cite{Frazier2009}. This section first reviews the case of approximating the knowledge-gradient in the presence of continuous parameters. Next an explicit formulation in case of deterministic observations is derived.

\subsection{Knowledge-Gradient for Continuous Parameters}
When $\mathcal{X}$ represents a continuous decision domain, \cref{eq:kgcb} can no longer be computed. A straightforward solution is discretizing $\mathcal{X}$ and handling the problem as a discrete decision problem. However, the computational complexity grows rapidly as the number of feasible decisions grows: this occurs when the dimensionality of the problem is not small, or the ranges of parameters are large. 

An approximation method referred to as the Knowledge-Gradient for Continuous Parameters (KGCP) was introduced \cite{Scott2011}, avoiding large-scale discretization of $\mathcal{X}$. Instead of maximizing over the entire decision domain, it was shown that including only the past and current sampling decisions is sufficient:
\begin{equation}
\bar{\nu}^{\mbox{KG},n}(\bfx) = \mathbb{E}\left[ \underset{i=0,..,n}{\max} ~ \mu^{n+1}(\bfx_i) |\bfx_n = \bfx\right] - \underset{i=0,..,n}{\max} ~ \mu^{n}(\bfx_i)|_{\bfx_n=\bfx}.
\label{eq:kgcp}
\end{equation}
Computing $\bar{\nu}^{\mbox{KG},n}$ is possible using a similar method as originally proposed for the knowledge-gradient for discrete optimization. The policy for optimal decision making, allowing $N$ observations of the problem, is summarized in \cref{alg:kgcp}. Sequentially the maximum of $\bar{\nu}^{\mbox{KG},n}$ is chosen as new decision and the model is improved with the acquired observation. At the end of the sampling process the final model is optimized and the location of global optimum $\overset{\star}{\bfx}$ is returned as optimal decision.

\begin{algorithm}[tb]
   \caption{KGCP Policy}
   \label{alg:kgcp}
\begin{algorithmic}
   \FOR{$n=0$ {\bfseries to} $N-1$}
   \STATE{$\bfx_n \in \underset{\bfx \in \mathcal{X}}{\argmax} ~ \bar{\nu}^{\mbox{KG},n}(\bfx)$}
   \STATE{Obtain observation $y_n$ for decision $\bfx_n$}
   \STATE{Calculate $\mu^{n+1}$ (or update $\mu^{n}$)}
   \ENDFOR
   \STATE Estimate of optimal decision $\overset{\star}{\bfx} \in \underset{\bfx \in \mathcal{X}}{\argmax} ~\mu^{N}(\bfx)$
\end{algorithmic}
\end{algorithm}

\subsection{Computing KGCP for deterministic problems}
\begin{figure*}[t]
\centering
\subfloat[]{\includegraphics[width=0.45\textwidth]{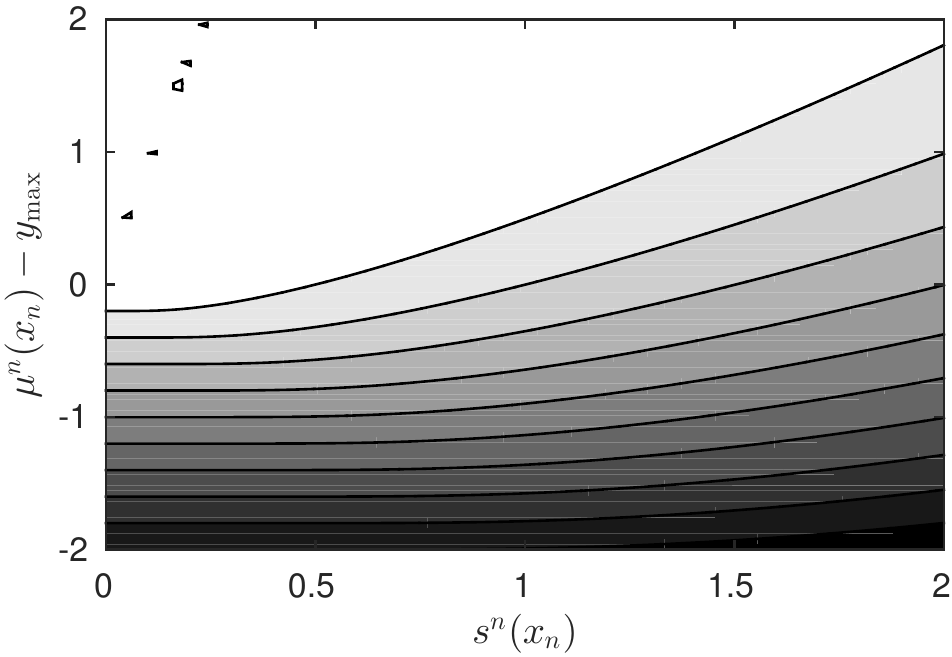}}%
~
\subfloat[]{\includegraphics[width=0.45\textwidth]{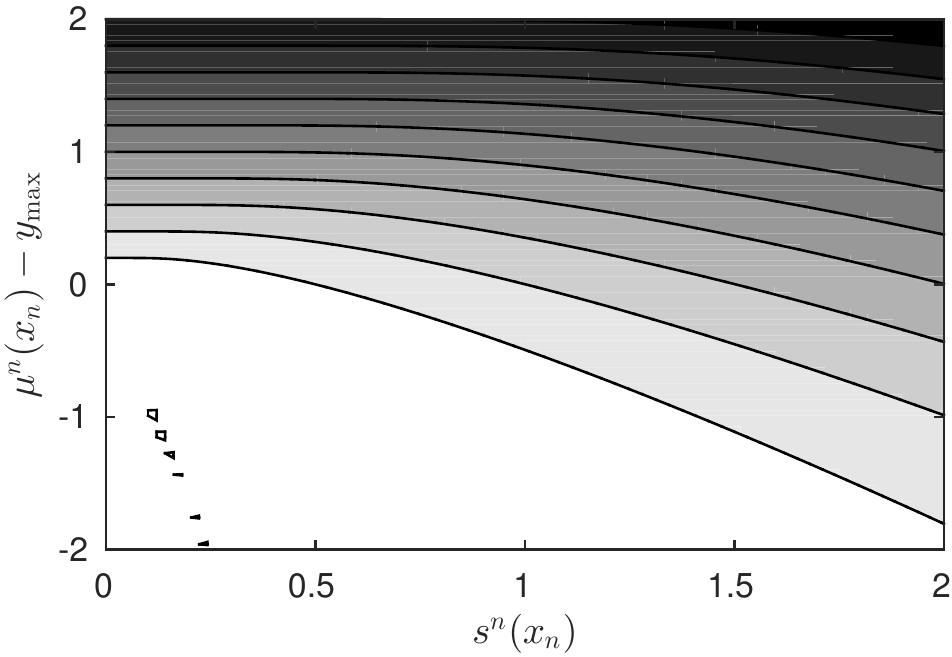}}%
\\
\subfloat[]{\includegraphics[width=0.45\textwidth]{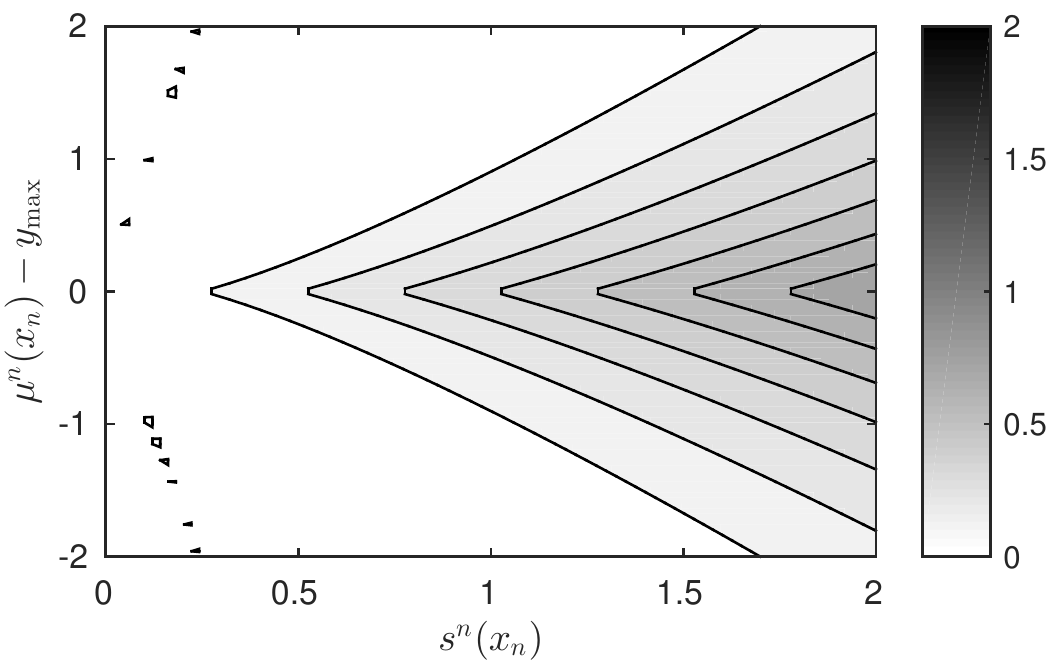}}%
\caption{Illustration of different policies as a function of prediction variance and improvement. a) Expected Improvement $\mathbb{E} \left[ I^n \right]$. b) Expected Decrement as defined in \cref{eq:eimin-outcome}. c) $\bar{\nu}^{\mbox{KG},n}$, as defined in \cref{eq:kgcp-simplified}. Clearly the latter is more conservative and only yields elevated scores if the prediction variance indicates improvement over $y_{\max}$ could occur.}
\label{fig:policies}
\end{figure*}

Denoting improvement for maximization problems as $I^n = \mu^{n+1}(\bfx) - y_{\max}$ with $y_{\max} = \max_{i=0,..,n-1} ~ y_i$, it was shown \cite{Scott2011} that
\begin{displaymath}
\bar{\nu}^{\mbox{KG},n} \leq \mathbb{E}\left[ I^n\right]
\end{displaymath} 
under the assumption that the observations are deterministic ($\lambda(\bfx_i) = 0$). The latter quantity corresponds to the Expected Improvement (EI) criterion \cite{Movckus1975, Jones1998} and is used widely as part of Efficient Global Optimization (EGO) in many applications in simulation optimization \cite{Couckuyt2010,Gazda2013,Mehari2015}. EI can be reformulated as follows:

\begin{align}
\mathbb{E}\left[ I^n\right] &= \mathbb{E}\left[ \max \left( \mu^{n+1}(\bfx)- y_{\max},0 \right)  \right] \nonumber \\
&=\mathbb{E}\left[ \max \left( \mu^{n+1}(\bfx),y_{\max} \right)  \right]-y_{\max} \nonumber\\
&=\mathbb{E}\left[ \max \left( \mu^{n+1}(\bfx),\underset{i=0,..,n-1}{\max} ~ \mu^{n}(\bfx_i) \right)  \right]-\underset{i=0,..,n-1}{\max} ~ \mu^{n}(\bfx_i) \label{eq:ei-deterministic-replacement}\\
&=\mathbb{E}\left[ \underset{i=0,..,n}{\max} ~ \mu^{n+1}(\bfx_i) |\bfx_n = \bfx\right]-\underset{i=0,..,n-1}{\max} ~ \mu^{n}(\bfx_i) \label{eq:ei-result}
\end{align}
In \cref{eq:ei-deterministic-replacement}, we used the property $y_i = \mu^n(\bfx_i) = \mu^{n+1}(\bfx_i), i=0,..,n-1$ which holds because of the deterministic assumption. The EI has an elegant closed form that is easy to compute. This form is usually written for minimization problems, for maximization the closed form is (the argument $\bfx_n$ is omitted for clarity):
\begin{align}
\label{eq:eimax}
\mathbb{E}\left[ I^n\right] &= \left(\mu^n-y_{\max}\right) \Phi \left( -z \right) + s^n \phi \left( z \right), \\
z &= \frac{y_{\max} - \mu^n}{s^n} \nonumber .
\end{align}
Comparing \cref{eq:kgcp} and \cref{eq:ei-result}, the difference is in the second term: the KGCP includes the model prediction for the current sampling decision whereas EI only includes all  previous observations. In order to define the exact relation between KGCP and EI, we infer the inequality case occurring when $\mu^{n}(\bfx) > y_{\max}$. Defining this case as the \textit{expected decrement}:

\begin{align}
\mathbb{E} \left[ D^n \right] &= \mathbb{E}\left[ \underset{i=0,..,n}{\max} ~ \mu^{n+1}(\bfx_i) |\bfx_n = \bfx\right]-\mu^{n}(\bfx) \nonumber\\
 &= \mathbb{E} \left[ \max \left(\mu^{n+1}(\bfx_n)-\mu^n(\bfx_n), y_{\max}-\mu^n(\bfx_n)\right)|\bfx_n=\bfx \right].\label{eq:beforeintegration}
\end{align}
By integrating out the expectation, we find an explicit formulation for this quantity as well.
\begin{proposition}
Under the assumption $\mu^{n+1}(\bfx_n) \sim \mathcal{N}\left( \mu^n(\bfx_n), s^n(\bfx_n) \right)$, the expected decrement corresponds to EI for minimization \label{eq:eimin-outcome}
\begin{align}
\mathbb{E} \left[ D^n \right] &= \left(y_{\max}-\mu^n\right) \Phi \left( z \right) + s^n \phi \left( z \right), \nonumber \\
&= \mathbb{E}\left[ -I^n \right] \nonumber.
\end{align}
\begin{proof}
The proof is given in \cref{sec:derivationed}.
\end{proof}
\end{proposition}
$\Phi$ and $\phi$ represent the standard normal cumulative and probability density functions respectively. We can now rewrite the KGCP for deterministic problems
\begin{equation}
\bar{\nu}^{\mbox{KG},n}(\bfx) = \min \left(\mathbb{E} \left[ I^n \right], \mathbb{E} \left[ D^n \right] \right). 
\label{eq:kgcp-simplified}
\end{equation} 

\begin{figure}
\centering
\subfloat[]{\includegraphics[width=0.44\columnwidth]{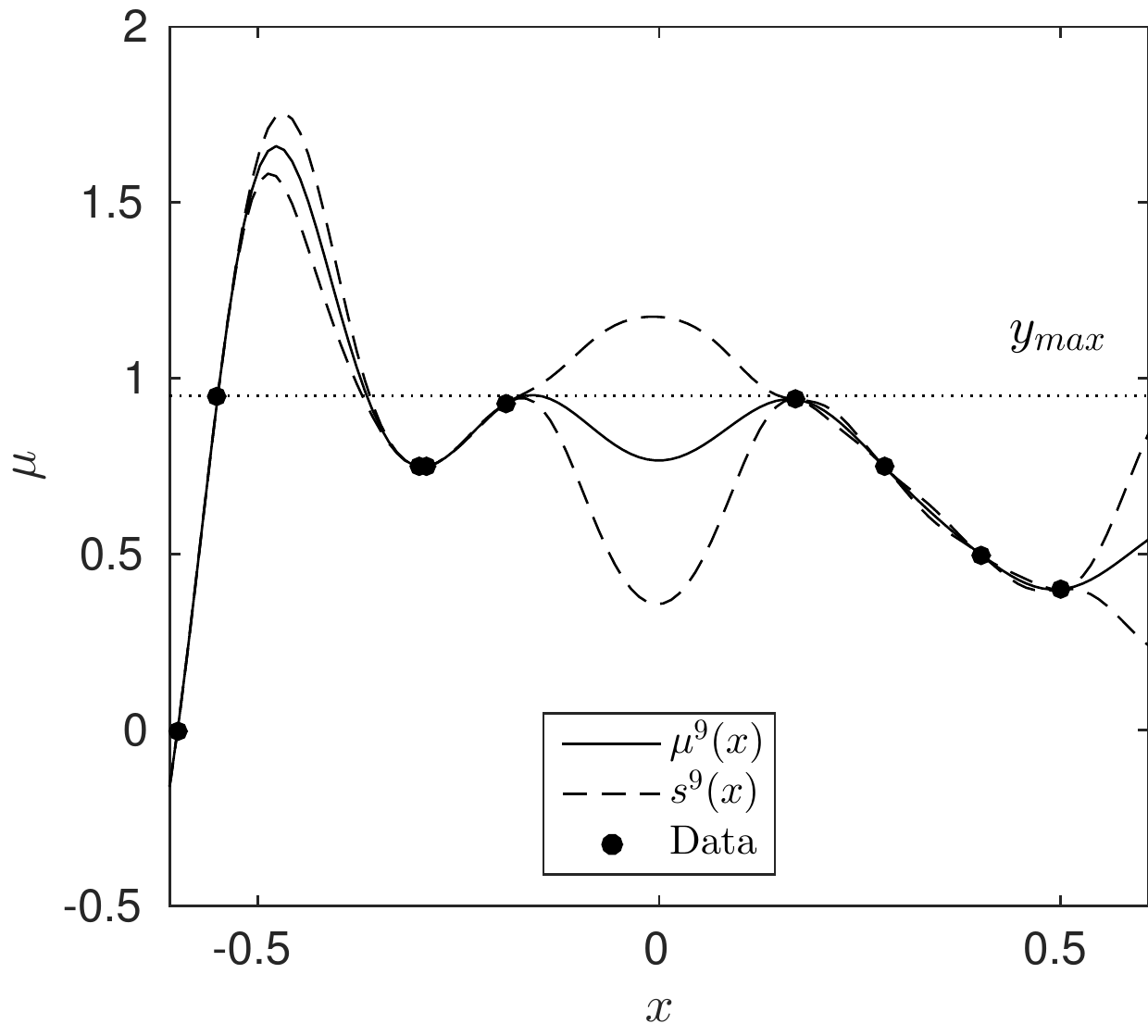}} 
~
\subfloat[]{\includegraphics[width=0.45\columnwidth]{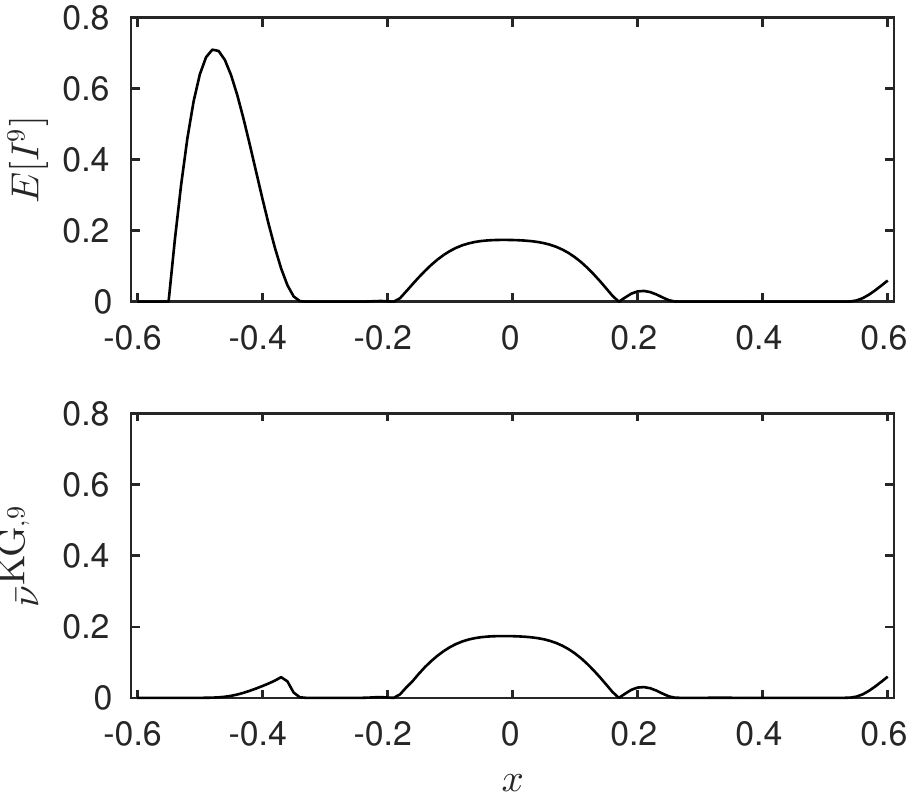}} 
\caption{Comparison of EI and KGCP criteria for 1D example. In (a) 9 decisions have been observed, and interpolated with a Kriging model. The prediction mean and variance are shown. In (b) the EI and KGCP policies for the same interval are shown, upper and bottom graph respectively. The leftmost area is ignored by KGCP because $\mathbb{E} \left[ D^n \right] < \mathbb{E} \left[ I^n \right]$ due to very low prediction variance. Instead, it focusses on the central region.}
\label{fig:difference}
\end{figure}

\Cref{fig:policies} illustrates the response of $\mathbb{E} \left[ I^n \right]$, $\mathbb{E} \left[ D^n \right]$ and $\bar{\nu}^{\mbox{KG},n}$ as a function of $\mu^{n}(\bfx) - y_{\max}$ and $s(\bfx)$. In the context of a maximization problem the expected decrement serves a similar purpose as EI for maximization: it indicates decisions which are believed to lead to worse results compared to the best decision we have observed ($y_{\max}$). The term tends to be smaller than $ \mathbb{E} \left[ I^n \right] $ in \cref{eq:kgcp-simplified} in regions satisfying $\mu^n(\bfx) > y_{\max}$ and small values for $s(\bfx)$. This corresponds to areas of which the model strongly expects improvement upon $y_{\max}$. Because of this notion of \textit{certain improvement} the quantity of $\mathbb{E} \left[ D^n \right]$ is smaller compared to $ \mathbb{E} \left[ I^n \right] $, hence KGCP ends exploitation and focusses on exploration instead.  

An illustration of a 1D maximisation problem is given in \cref{fig:difference}: two areas have elevated EI scores; however the most promising area according to EI is assigned a lower KGCP score: although the model indicates the objective function can be improved, $\mathbb{E} \left[ D^n \right]$ is very small because of a low variance. The KGCP policy trusts the belief of the model and does not further verify this area. The KGCP policy decides to explore the center area instead as it is more uncertain and could contain a new optimum. Note that the prediction variance of the model is more important for KGCP than EI: this could imply KGCP benefits more from slice sampling. We test this hypothesis empirically in \cref{sec:exp}.

\subsection{Gradient of Deterministic KGCP}
To facilitate the optimization problem as defined in \cref{eq:kgcpoptimization}, we investigate the computation of the gradient of \cref{eq:kgcp-simplified} as required by optimization methods such as conjugate gradient \cite{Gill1979}. The $\mathbb{E} \left[ I^n \right]$ is a differentiable function, it can be computed as follows:
\begin{subequations}
\begin{align}
\frac{\dif \mathbb{E} \left[ I^n \right]}{\dif \bfx} &= \left( -z \Phi(-z) + \phi (z) \right) \frac{\dif s^n}{\dif \bfx}- s^n \Phi (-z) \frac{\dif z}{\dif \bfx} \label{eq:deriv-ei}\\
\frac{\dif \mathbb{E} \left[ D^n \right]}{\dif \bfx} &= \left( z \Phi(z) + \phi (z) \right) \frac{\dif s^n}{\dif \bfx}+ s^n \Phi (z) \frac{\dif z}{\dif \bfx} \label{eq:deriv-ed}\\
\frac{\dif z}{\dif \bfx} &= -\frac{\frac{\dif \mu^n}{\dif \bfx} + z \frac{\dif s^n}{\dif \bfx}}{s^n}
\end{align}
\end{subequations}
However, the $\min$ function in \cref{eq:kgcp-simplified} complicates defining the gradient. The approach presented in \cite{Zhang1996} can be used to compute a gradient however it is not guaranteed to exist everwhere. 
\begin{lemma}
\label{lemma:equalityofcriteria}
\begin{displaymath}
\forall \bfx \in \mathcal{X}: ~ \mathbb{E} \left[ I^n \right] = \mathbb{E} \left[ D^n \right] \Longleftrightarrow y_{\max} = \mu^n(\bfx)
\end{displaymath}
\begin{proof}
\begin{align*}
 & &\mathbb{E} \left[ I^n \right] &= \mathbb{E} \left[ D^n \right]& \\
\Leftrightarrow & &(\mu^n - y_{\max}) \Phi(-z) &= (y_{\max} - \mu^n) \Phi(z)& \\
\Leftrightarrow & &\mu^n &=y_{\max}&
\end{align*}
\end{proof}
\end{lemma}
The result of \cref{lemma:equalityofcriteria} implies $\mathbb{E} \left[ I^n \right]$ equals $\mathbb{E} \left[ D^n \right]$ when $z=0$. Unfortunately, \cref{eq:deriv-ei} and \cref{eq:deriv-ed} are not equal for  this case, hence the gradient of \cref{eq:kgcp-simplified} is not guaranteed to exist over the entire search domain \cite{Zhang1996}.

Fortunately, both quantities as required by \cref{eq:kgcp-simplified} can be computed very efficiently which means derivative-free meta-heuristics such as Particle Swarm Optimization (PSO) \cite{Kennedy2010} can be used to maximize the KGCP. Should a derivative be required, the minimum function must be replaced by a smoother alternative. The minimum function corresponds to the $l_{-\infty}$ norm, hence it can be approximated by any $l_{a}$ with $a \in \mathbb{Z}^{-}, a \ll 0$ which results in a differentiable form, or by applying a soft minimum of the followin soft form of the KGCP for deterministic problems:
\begin{subequations}
\begin{align}
\bar{\nu}^{\mbox{KG},n}_{s}(\bfx) &= -\frac{\log \left( \exp \left(-k\mathbb{E} \left[ I^n \right] \right) + \exp \left( -k\mathbb{E} \left[ D^n \right] \right) \right)}{k} \\
\frac{\dif \bar{\nu}^{\mbox{KG},n}_{s}(\bfx)}{\dif \bfx} &= \frac{\exp \left(k\mathbb{E} \left[ I^n \right]\right)\frac{\dif \mathbb{E} \left[ D^n \right]}{\dif \bfx} + \exp \left(k\mathbb{E} \left[ D^n \right]\right)\frac{\dif \mathbb{E} \left[ I^n \right]}{\dif \bfx}}{\exp \left(k\mathbb{E} \left[ I^n \right] \right) + \exp \left( k \mathbb{E} \left[ D^n \right] \right)}.
\label{eq:kgcp-smooth}
\end{align} 
\end{subequations}
The constant $k > 0$ controls the smoothing, for $k=\infty$ the soft version of the KGCP is equivalent to the hard version of \cref{eq:kgcp-simplified}.
 
\section{Experiments}
\label{sec:exp}
We implemented the KGCP formulation of \cref{eq:kgcp-simplified} in the SUMO-Toolbox \cite{Gorissen2010, Vanderherten2016}, a research platform for surrogate modeling supporting grid-based computing for global surrogate modeling and simulation optimization. The toolbox is easily extendable and already contains robust and tested implementations of Kriging and Expected Improvement for comparison purposes \cite{Couckuyt2014}. It also ships an implementation of Upper Confidence Bound (UCB) \cite{Freitas2012}, another sampling policy for optimization problems, which is included as third method throughout our tests.

\subsection{Test setup}
We compare six different test setups: three different policies (KGCP, EI and UCB) for selecting the next decision are combined with two different methods (MLE and slice sampling as described in Section~\ref{sec:hpoptim}) for determining the hyperparameters $\theta$ of the correlation function. This results in the following test configurations: \texttt{KGCP-MLE}, \texttt{KGCP-SS}, \texttt{EI-MLE}, \texttt{EI-SS}, \texttt{UCB-MLE} and \texttt{UCB-SS}. For the slice sampling, 100 settings of $\theta$ are sampled from the marginal likelihood: the MLE estimate is used as starting point. The test setups are applied to several optimization test problems as described in the following sections. Each test setup was replicated 100 times on each test problem as variance in the results is expected: particularly on the multimodal test problems as these rely strongly on exploration for discovery of the optima. The results are averaged, and 95\% confidence intervals are computed. 

As a starting point, the test problem is first evaluated on a (maximin) Latin Hypercube of size 10, generated by the Translational Propagation algorithm \cite{Viana2010}. After obtaining the corresponding observations, a model is built and its hyperparameters are determined. From there on the sampling policy sequentially identifies a new decision: the policy is optimized by first applying Monte Carlo sampling, and optimizing the 10 best candidates using local search. When a choice is made for the decision $\bfx$, the observation is obtained and a new model is trained. This process continues until a pre-set number of $N$ observations have been obtained. 

Each iteration $n$, the progress of finding the global optimum is evaluated by computing the Opportunity Cost (OC). Defining $\overset{\star}{i} = \argmax_i ~\mu^{n}(i)$ for the intermediate model $\mu^n$, the OC equals:
\begin{equation}
\mbox{OC} = \underset{i}{\max} ~ \mu(i) - \mu(\overset{\star}{i}).
\end{equation} 
The OC represents how close the solution has come to the global optimum, if the process were to be ended. Note that usually in EGO applications $\overset{\star}{i} = y_{\max}$. Should the model believe some areas have a model response which is better than $y_{\max}$ it is usually ignored. The KGCP policy however aims to select decisions to optimally improve the belief of the model so it can identify the location of the best decision. Hence $\overset{\star}{i}$ is chosen as the optimum of the model.

\subsection{Synthetic test problems}
\begin{figure}
		\centering
        \subfloat[][Branin]{
        		\centering
                \includegraphics[width=0.45\textwidth]{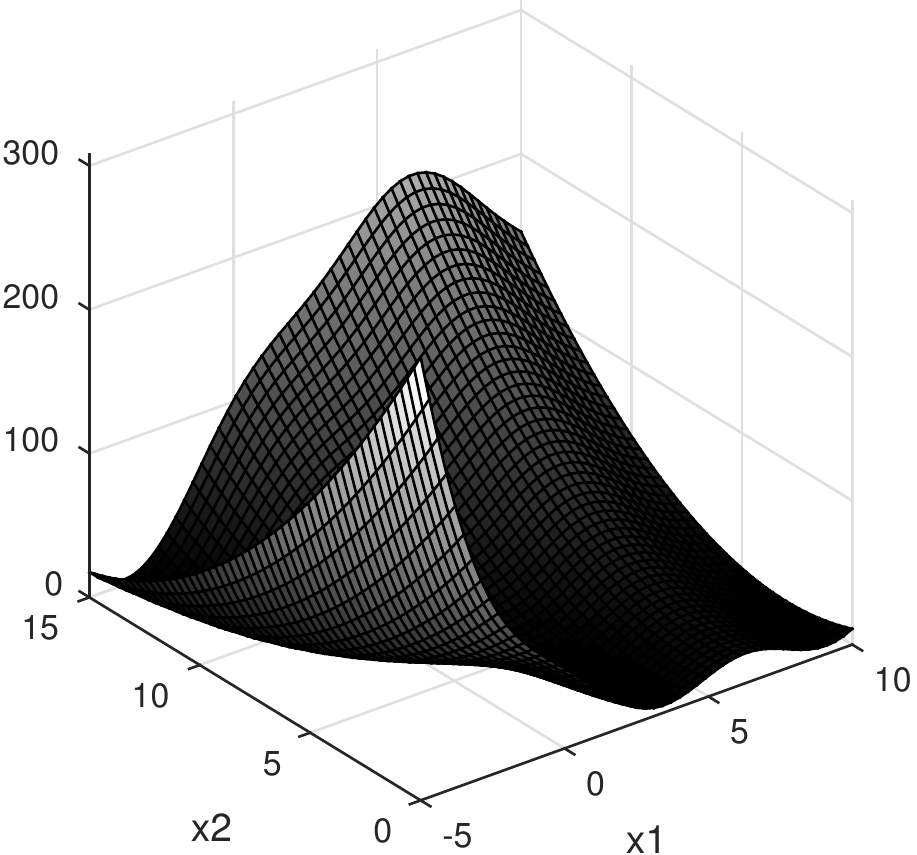}
                \label{fig:ld-branin}
        }
        ~ 
        \subfloat[][Hartmann (slice near optimum)]{
        		\centering
                \includegraphics[width=0.45\textwidth]{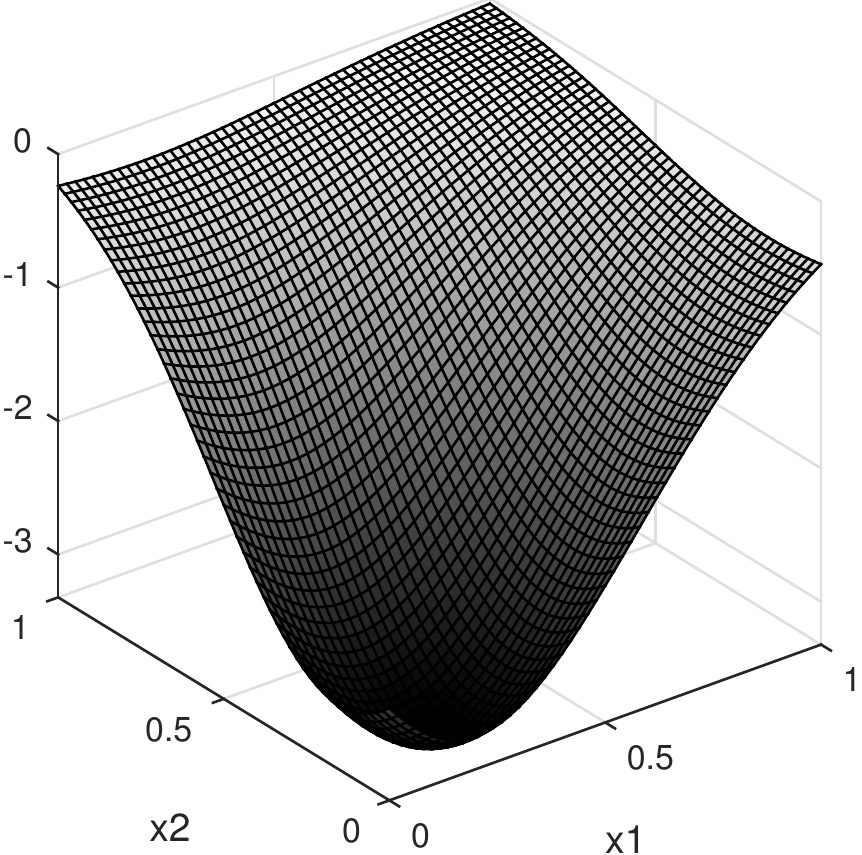}
                \label{fig:ld-hartmann}
        }
        \\
        \subfloat[][Schwefel]{
        		\centering
                \includegraphics[width=0.45\textwidth]{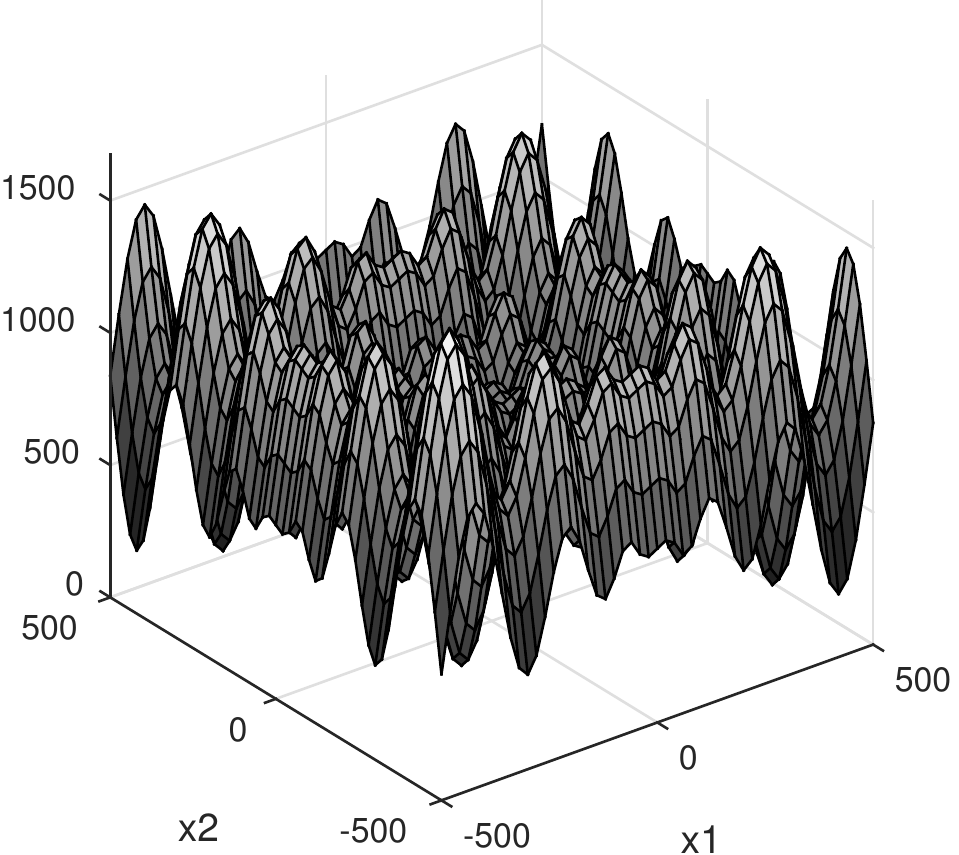}
                \label{fig:ld-schwef}
        }
        ~
        \subfloat[][Eggholder]{
        		\centering
                \includegraphics[width=0.45\textwidth]{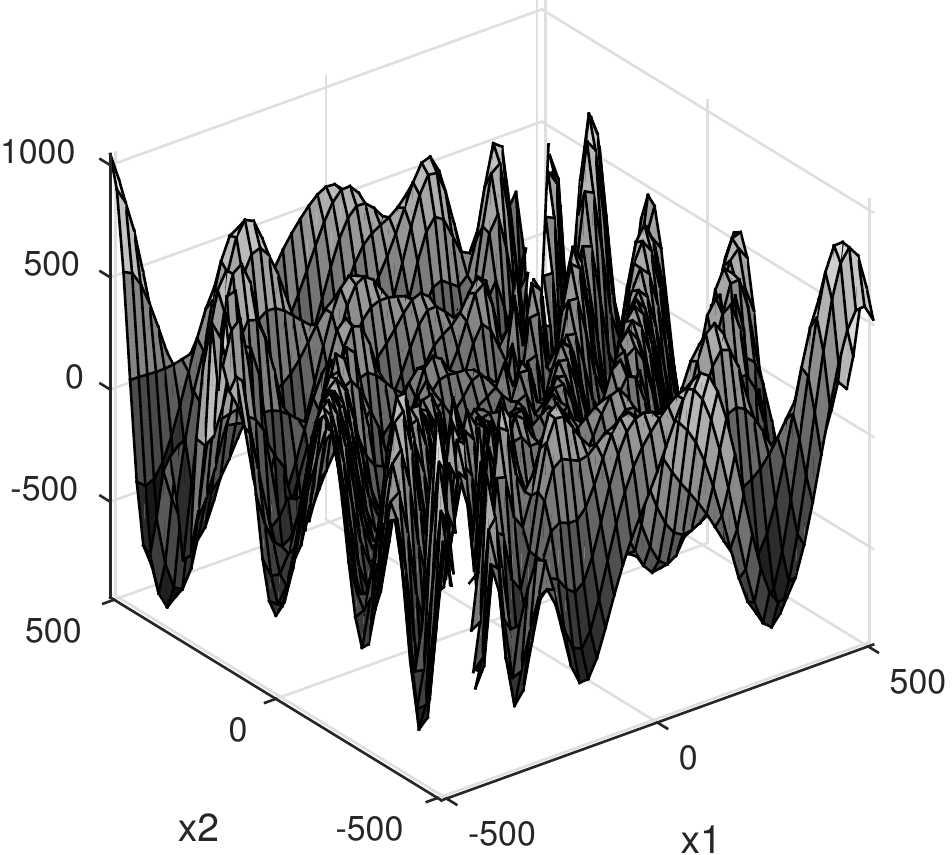}
                \label{fig:ld-egg}
        }
        \caption{Illustrations of the synthetic test cases}
        \label{fig:testillustrations}
\end{figure}

First, the test setups are applied to four different synthetic functions for global optimization. All of them are defined for minimization so in our experiments, functions are negated. We included both easy and harder test functions, as well as a 6D test problem. All functions are illustrated in \cref{fig:testillustrations}.

\subsubsection{Case 1: Branin 2D}
The Branin function \cite{Forrester2008b} is used often as an optimization benchmark function. It has two input parameters with ranges $[-5, 10]$ and $[0, 15]$, respectively, and three optima. The response surface is very smooth and does not contain any abrupt discontinuities. Optimization of this function is straightforward, and in total $N = 20$ points are evaluated.  

\subsubsection{Case 2: Hartmann 6D}
A second test problem considered is the Hartmann function \cite{Picheny2013} with 6 input parameters on the interval $[0,1]$. The function has a few local optima but is not very difficult to optimize as the response surface is again quite smooth. Though the size of its search space is considerably larger due its dimensionality. The total number of evaluations was restricted to $N = 40$.

\subsubsection{Case 3: Schwefel 2D}
The Schwefel function \cite{Laguna2005} is a more challenging problem compared to Branin and Hartmann as it contains many optima. The response surface is quite bumpy but still easy to characterize using Kriging, in contrast to the next test problem (Eggholder). The range considered was $[-500,500]$, for both input parameters. The actual global optimum is situated at $(420.9687, 420.9687)$. Because the optimization problem is more difficult, we set $N = 100$.

\subsubsection{Case 4: Eggholder 2D}
The last synthetic test problem is the \\ Eggholder function \cite{Jamil2013}. This function is considered very difficult to optimize because it contains large numbers of local optimum and has a very tough surface to characterize. It has two input parameters over the range $[-512, 512]$ and the global optimum is situated at $(512, 404.2319)$. It is surrounded by several local optima which tend to trap optimization algorithms as they are very deceptive (because they are quite steep). Like the Schwefel function the optimization of this function was given a computational budget of $N = 100$.

\subsubsection{Results}
\begin{table}
\caption{Mean OC and 95\% Confidence Intervals (CI) obtained after obtaining $N$ observations by averaging over 100 runs, for each test setup for determining the hyperparameters $\bftheta$ applied to the test problems.}
\label{tab:results}
\centering
\subfloat[KGCP]{
\begin{tabular}{ |l c || c c | c c|}
\hline
& & \multicolumn{2}{c|}{MLE} & \multicolumn{2}{c|}{SS} \\
\hline
Problem & $N$ & $\mathbb{E}\left( \text{OC} \right)$ & 95\% CI & $\mathbb{E}\left( \text{OC} \right)$ & 95\% CI \\
\hline
\specialcell[l]{Branin \\ $d = 2$} & 20 & 0.006 & \numrange{0.009}{0.003} & 0.025 & \numrange{0}{0.06}\\
\hline
\specialcell[l]{\textbf{Hartmann} \\ $d = 6$} & 40 & \textbf{2.12} & \numrange{2.12}{2.13} & 2.14 & \numrange{2.13}{2.15}\\
\hline
\specialcell[l]{\textbf{Schwefel} \\ $d = 2$} & 100 & \textbf{124.0} & \numrange{114.8}{133.1} & 156.2 & \numrange{150.7}{179.7}\\
\hline
\specialcell[l]{Eggholder \\ $d = 2$} & 100 & 48.0 & \numrange{42.8}{53.2} & 60 & \numrange{41.8}{78.1}\\
\hline
\hline
\specialcell{\textbf{Truss} \\ $d = 10$} & 250 & 1.64 & \numrange{1.44}{1.84} & \textbf{1.17} & \numrange{1.0}{1.35}\\
\hline
\end{tabular}
} \\
\subfloat[EI]{
\begin{tabular}{ |l c || c c | c c|}
\hline
& & \multicolumn{2}{c|}{MLE} & \multicolumn{2}{c|}{SS} \\
\hline
Problem & $N$ & $\mathbb{E}\left( \text{OC} \right)$ & 95\% CI & $\mathbb{E}\left( \text{OC} \right)$ & 95\% CI \\
\hline
\specialcell[l]{Branin \\ $d = 2$} & 20 & 0.008 & \numrange{0.001}{0.014} & 0.008 & \numrange{0.005}{0.010}\\
\hline
\specialcell[l]{Hartmann \\ $d = 6$} & 40 & 2.13 & \numrange{2.13}{2.14} & 2.13 & \numrange{2.13}{2.13}\\
\hline
\specialcell[l]{Schwefel \\ $d = 2$} & 100 & 151.2 & \numrange{140.8}{161.7} & 154.0 & \numrange{139.4}{168.5}\\
\hline
\specialcell[l]{\textbf{Eggholder} \\ $d = 2$} & 100 & 81.2 & \numrange{65.0}{97.3} & \textbf{46.41} & \numrange{36.3}{56.5}\\
\hline
\hline
\specialcell{Truss \\ $d = 10$} & 250 & 1.76 & \numrange{1.59}{1.92} & 1.50 & \numrange{1.31}{1.70}\\
\hline
\end{tabular}
} \\
\subfloat[UCB]{
\begin{tabular}{ |l c || c c | c c|}
\hline
& & \multicolumn{2}{c|}{MLE} & \multicolumn{2}{c|}{SS} \\
\hline
Problem & $N$ & $\mathbb{E}\left( \text{OC} \right)$ & 95\% CI & $\mathbb{E}\left( \text{OC} \right)$ & 95\% CI \\
\hline
\specialcell[l]{\textbf{Branin} \\ $d = 2$} & 20 & \textbf{0} & \numrange{0}{0} & \textbf{0} & \numrange{0}{0}\\
\hline
\specialcell[l]{Hartmann \\ $d = 6$} & 40 & 2.13 & \numrange{2.12}{2.13} & 2.13 & \numrange{2.12}{2.13}\\
\hline
\specialcell[l]{Schwefel \\ $d = 2$} & 100 & 236.9 & \numrange{236.9}{236.9} & 233.1 & \numrange{233.1}{233.1}\\
\hline
\specialcell[l]{Eggholder \\ $d = 2$} & 100 & 143.3 & \numrange{125.7}{160.9} & 149.8 & \numrange{132.2}{167.4}\\
\hline
\hline
\specialcell{Truss \\ $d = 10$} & 250 & 4.63 & \numrange{4.19}{5.06} & 2.39 & \numrange{2.02}{2.76}\\
\hline
\end{tabular}
}
\end{table}

\begin{figure*}
\centering
\subfloat[]{\includegraphics[width=0.49\textwidth]{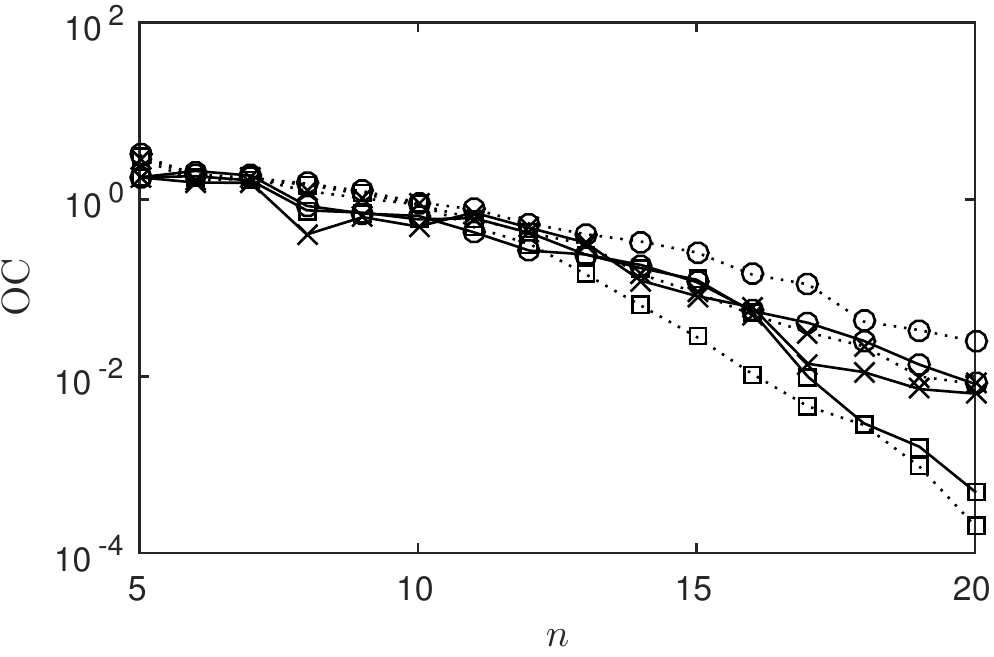}}%
~
\subfloat[]{\includegraphics[width=0.49\textwidth]{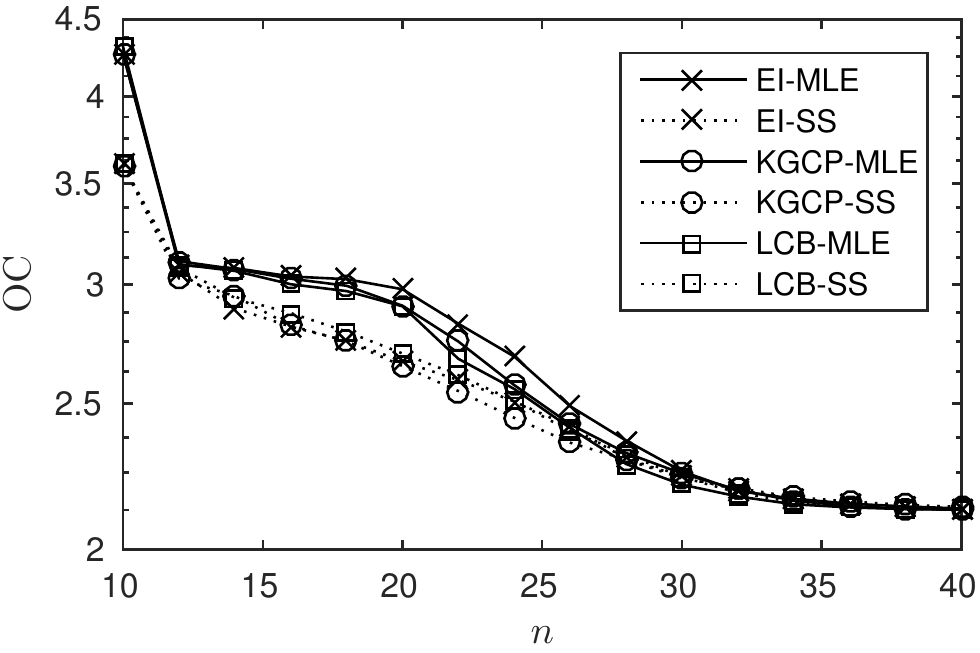}}%
\\
\subfloat[]{\includegraphics[width=0.49\textwidth]{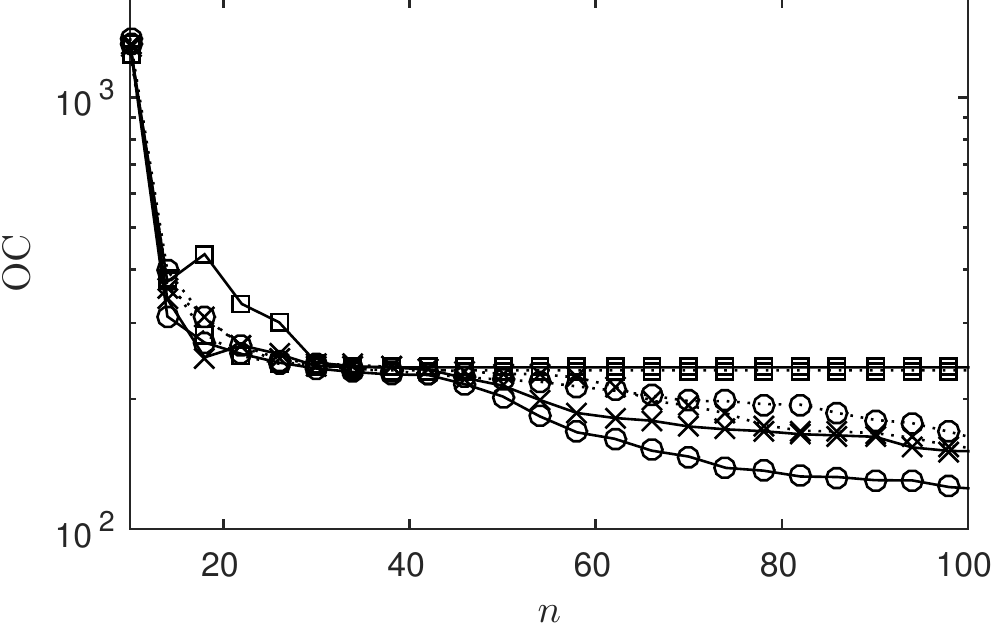}}%
~
\subfloat[]{\includegraphics[width=0.49\textwidth]{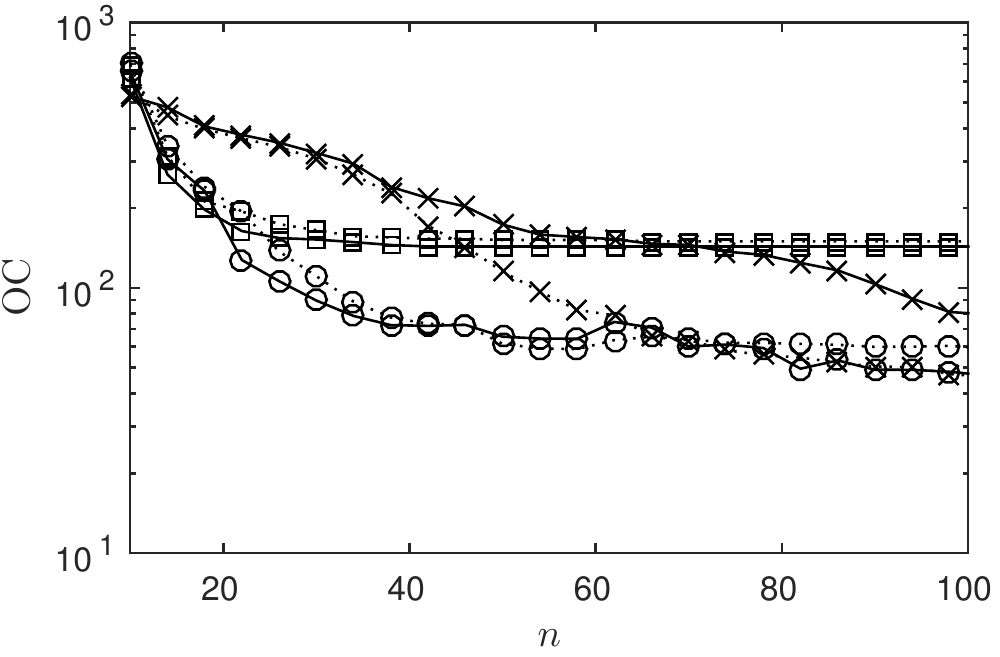}}%
\caption{Synthetic test problems: evolution of the mean OC as more observations are obtained. a) Branin 2D, b) Hartmann 6D, c) Schwefel 2D, d) Eggholder 2D.}
\label{fig:synthetic-resuls-plots}
\end{figure*}

The results for all experiments are summarized in Table~\ref{tab:results}. For all test setups the mean OC and its $95$\% Confidence Intervals (CI) are shown at the end of the runs (after $N$ evaluations). In addition, the evolution of the OC as more observations are obtained for the synthetic test problems is shown in Figure~\ref{fig:synthetic-resuls-plots}.

The Branin function does not require many evaluations in order to find the global optimum. All test configurations perform similar except \texttt{EI-MLE} which is stuck in a poor choice of hyperparameters for a few iterations: the optimum of the model however is near the true optimum. As more observations are available the situation is corrected causing a rising OC. Note that \texttt{EI-SS}, \texttt{KGCP-MLE} and \texttt{KGCP-SS} avoid this situation. However, for this problem \texttt{UCB-MLE} and \texttt{UCB-SS} find better solutions. The UCB method orients more towards exploitation, which works well for this test problem as the true optimum is not difficult to find.

The Schwefel test problem however has a more difficult response surface. A small difference appears: \texttt{KGCP-MLE} reduces the OC faster. When about 50 observations have been obtained the runs clearly obtain a lower OC. The other test setups hardly differ, with the exception of the UCB cases which quickly get stuck in a local minimum. Note that the \texttt{KGCP-SS} setup performs worse in comparison to both EI setups. 

For the most difficult synthetic test problem, the Eggholder function, the difference is more significant. Again, \texttt{UCB-MLE} and \texttt{UCB-SS} both quickly get stuck in a local minimum. For this test problem, the \texttt{KGCP-MLE} and \texttt{KGCP-SS} configurations quickly improve a lot more compared to \texttt{EI-MLE} and \texttt{EI-SS}. The \texttt{EI-SS} setup performs better for this problem compared to \texttt{EI-MLE}, but only catches up with the KGCP setups after 20 additional observations. When 100 observations have been made, all test setups provide similar OC scores which means no clear winner is shown in Table~\ref{tab:results}, however KGCP runs for this test problems could already have ended after 40 observations and provide satisfactory results.

Generally, for the synthetic test problems \texttt{KGCP-MLE} obtains the best results. The other test setups are very comparable and show not much differences, with the exception of the UCB setups which are only satisfactory for the easier Branin and Hartmann test problems. The complexity of the Eggholder problem results in a significant difference between the EI and KGCP sampling policies. Surprisingly, we do not see an improvement of \texttt{KGCP-SS} over \texttt{KGCP-MLE}: in fact it performs worse in 3 out of 4 problems which is in contrast with our earlier hypothesis. A possible explanation to this is the nature of the synthetic problems: these can be interpolated nicely which implies a well-defined unimodal optimum of the likelihood that can be easily identified with MLE. Slice sampling draws most additional $\theta$ vectors near this optimum of the MLE estimate but the corresponding models are likely to be affected negatively. Hence errors are averaged and the policy is not performing as powerful. By lowering the stepsize of the slice sampling, results are expected to be more in line with those obtained by MLE, however one may argue that the additional computation is no longer worth it. In other words, for these synthetic functions there is not a lot of uncertainty on the parameter estimate, and the MLE solution on its own appears to be most powerful. 

\subsection{Truss structure optimization (10D)}
\begin{figure}[t]
\centering
\includegraphics[width=0.8\columnwidth]{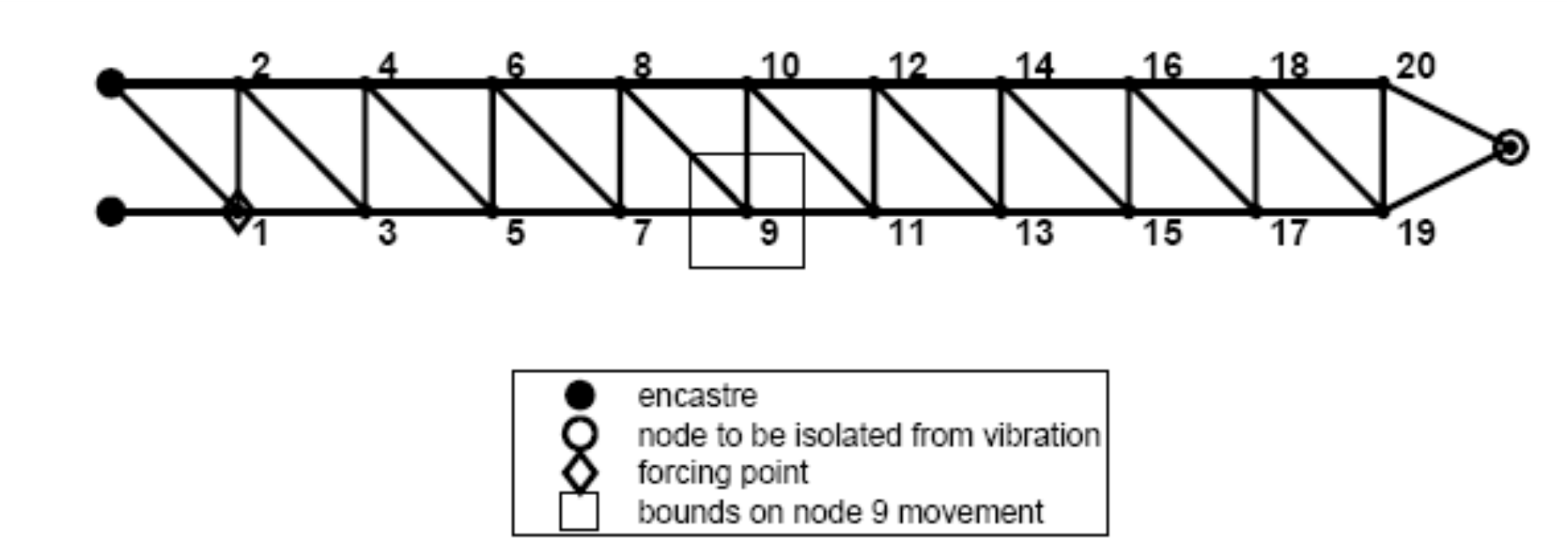}%
\caption{The two-dimensional truss structure (courtesy of \cite{Forrester2008})}
\label{fig:truss}
\end{figure}

Our final test problem is a structural dynamics problem of a two-dimensional truss for maximum passive vibration isolation. The truss is constructed using 42 Euler-Bernoulli beams having two finite elements per beam. The truss is subjected to unit force excitation at node 1, across the \SIrange{100}{200}{\hertz} frequency range. The input parameters correspond to moving 5 nodes (8 through 12) in a $0.9 \times 0.9$ square. The other nodes are kept fixed according to the structure depicted in Figure~\ref{fig:truss}. The band-averaged vibration attenuation at the tip, compared to the baseline structure is to be maximized. The more nodes are included as design parameters, the more multi-modal the response is as many (sub-)optimal configurations are possible. It was reported earlier that standard EI tends to be tricked and has a low probability of identifying the global optimum \cite{Forrester2008}. Because this experiment is a difficult 10D problem, a total amount of $N = 250$ observations are allowed. The same sampling policies, both tested with the MLE and slice sampling approaches are tested, and each configuration is replicated 100 times to compute the 95\% confidence intervals.

\begin{figure}[t]
\centering
\includegraphics[width=0.7\columnwidth]{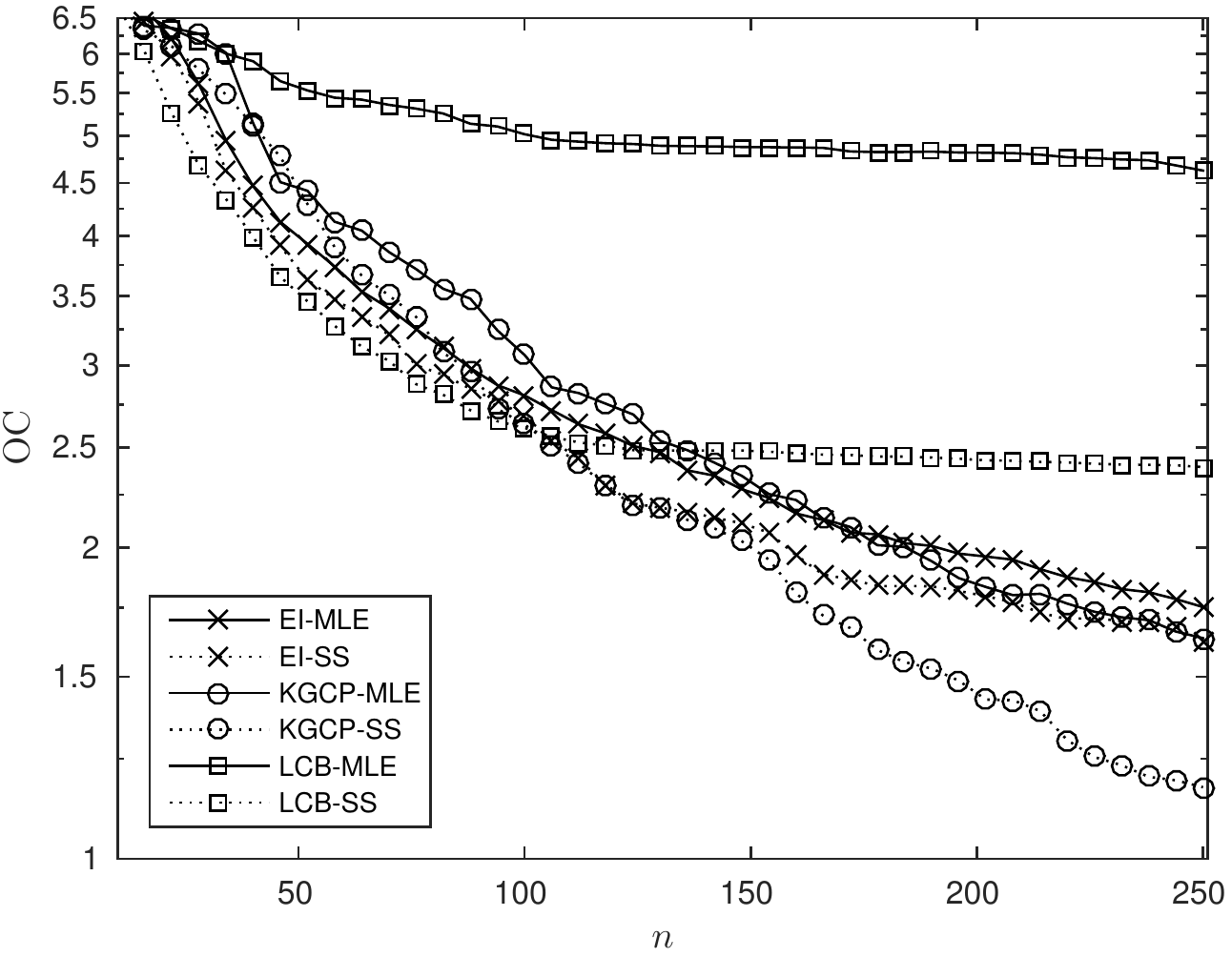}%
\caption{Truss problem (10D): evolution of the mean OC as more observations are obtained.}
\label{fig:truss-error}
\end{figure}

The results of running the test setups on this test problem are shown in Table~\ref{tab:results}, and a more detailed evolution of the OC as more observations are obtained is illustrated in Figure~\ref{fig:truss-error}. As opposed to the synthetic runs, the benefit of slice sampling starts to show on this structural optimization problem. The uncertainty on the point-estimate of the hyperparameters is larger and the inclusion of different hyperparameters clearly results in an improvement. Both \texttt{EI-MLE} and \texttt{KGCP-MLE} are less performant, the latter slightly beating the former. The best results are obtained by \texttt{KGCP-SS} as it keeps improving when 170 observations have been obtained, whereas \texttt{EI-SS} seems to stabilize at this point. However, at the beginning of the run the drop of the OC is slightly better for the \texttt{EI-SS} compared to the other test setups. Finally, the \texttt{UCB-MLE} and \texttt{UCB-SS} runs are performing significantly worse and seem to be stuck at some point. Using slice-sampling the result is slightly better but it is still outperformed by all other test setups.

\section{Conclusion}
In this work we derived a closed formula for fast computation of the KGCP for expensive optimization problems, assuming the response is noiseless (deterministic). We lifted the computational complexity disadvantage with respect to EI making the KGCP a feasible choice for this class of problems. For simple problems the KGCP formulation gives comparable results to the popular EI, but as the complexity increases (multimodal problems) the extra tendency of KGCP towards exploration helps to avoid being stuck for some time in a local optimum. Because the KGCP potentially gives better results faster and has similar complexity as EI, we suggest its usage.

It should be kept in mind that KGCP essentially trusts the model if it is certain a region contains a better optimum (better prediction mean, low prediction variance). Unfortunately this trust comes with a risk: it is crucial to ensure that the model, and more specifically the prediction variance is as accurate as possible. However if the optimum of the likelihood is well-defined and the surface is unimodal, only little uncertainty is expected on the model parameters, resulting in the configurations with MLE outperforming slice sampling in our synthetic experiments. For the structural optimization problem however, this is no longer the case. Hence the choice of whether to use slice sampling or not should be made on the type of problem, and more specifically the uncertainty expected on the hyperparameters which is indicated by the shape of their distribution (i.e., the likelihood), or in a more general sense, by the complexity of approximating the response.

\section*{Acknowledgements}
Experiments were carried out using the STEVIN Supercomputer Infrastructure at Ghent University, funded by Ghent University, the Flemish Supercomputer Center (VSC), the Hercules Foundation and the Flemish Government – department EWI. The authors would like to thank Alex. J. Forrester from Southampton University, UK, for providing the code for the passive vibration isolating truss.

\appendix
\section{Derivation of Expected Decrement}
Computation of the closed form of $\mathbb{E} \left[ D^n \right]$ is possible by integrating the expectation in \cref{eq:beforeintegration}. The maximization within the expectation is handled by splitting the integration range. Let $Y = \mu^{n+1}(\bfx_n) \sim \mathcal{N}\left( \mu^n(\bfx_n), s^n(\bfx_n) \right)$ (omitting $\bfx_n$ for presentation clarity):
\begin{align*}
\mathbb{E} \left[ D^n \right] &= \int_{-\infty}^{y_{\max}} \left( y_{\max}-\mu^n \right) \phi \left( Y |\mu^n, s^n \right) \dif Y + \int^{\infty}_{y_{\max}} \left( Y-\mu^n \right) \phi \left( Y |\mu^n, s^n\right) \dif Y  \\
&= \int_{-\infty}^{y_{\max}} \left( y_{\max}-\mu^n \right) \frac{\phi \left( \frac{Y-\mu^n}{s^n} \right)}{s^n} \dif Y + \int^{\infty}_{y_{\max}} \left( Y-\mu^n \right) \frac{\phi \left( \frac{Y-\mu^n}{s^n} \right)}{s^n} \dif Y
\end{align*}
Substituting $u = \frac{Y-\mu^n}{s^n}$:
\begin{align*}
&= \left( y_{\max}-\mu^n \right) \left[ \Phi \left( \frac{Y-\mu^n}{s^n}\right) \right]^{y_{\max}}_{-\infty} + \int_{A}^{\infty} (s^n u + \mu^n - \mu^n) \phi(u) \dif u \\
&= \left( y_{\max}-\mu^n \right) \Phi \left( \frac{y_{\max}-\mu^n}{s^n}\right) + s \int_{A}^{\infty} u \phi(u) \dif u \\
&= \left( y_{\max}-\mu^n \right) \Phi \left( \frac{y_{\max}-\mu^n}{s^n}\right) + s \left[ -\phi(u)\right]^\infty_A \\
&= \left( y_{\max}-\mu^n \right) \Phi \left( \frac{y_{\max}-\mu^n}{s^n}\right) + s\phi(A) \\
&= \left( y_{\max}-\mu^n \right) \Phi \left( \frac{y_{\max}-\mu^n}{s^n}\right) + s\phi\left(\frac{y_{\max}-\mu^n}{s^n}\right)
\end{align*}
\label{sec:derivationed}

\bibliographystyle{siamplain}
\bibliography{bibliography}

\end{document}